\documentclass[%
 reprint,
 amsmath,amssymb,
 aps,
]{revtex4-1}

\usepackage{graphicx}
\usepackage{dcolumn}
\usepackage{bm}

\begin{document}


\title{Wave-Particle duality in Single-Photon Entanglement}

\author{Wei Li$^{1,2,3}$}
\author{Shengmei Zhao$^{1,2}$}%
 \email{zhaosm@njupt.edu.cn}
\affiliation{$^{1}$Nanjing University of Posts and Telecommunications, Institute of Signal Processing and Transmission, Nanjing, 210003, China.}%
\affiliation{$^{2}$Nanjing University of Posts and Telecommunications, Key Lab Broadband Wireless Communication and Sensor Network, Ministy of Education, Nanjing, 210003, China.}%
\affiliation{$^{3}$National Laboratory of Solid State Microstructures, Nanjing University, Nanjing 210093, China.}%


\date{\today}

\begin{abstract}
The simplest single-photon entanglement is the entanglement of the vacuum state and the single-photon state between two path modes. The verification of the existence of single-photon entanglement has attracted extensive research interests. Here, based on the construction of Bell's inequality in wave space, we propose a new method to verify single photon entanglement. Meanwhile, we define the wave state in two-dimensional space relative to the photon number state, and propose a method to measure it. Strong violation of Bell inequality based on joint measurement of wave states indicates the existence of single photon entanglement with certainty. Wave state entanglement obtained from Fourier transform of single photon entanglement and the corresponding measurement protocols will provide us with more information-carrying schemes in the field of quantum information. The difference in the representation in photon-number space and wave space implies the wave-particle duality of single photon entanglement.
\end{abstract}

\pacs{Valid PACS appear here}
\maketitle

\section{\label{sec:level1}Introduction}

\par  It is generally believed that the number of particles in the system where quantum entanglement occurs is greater than or equal to 2\cite{pan2012multiphoton,horodecki2009quantum}. These particles entangled at some specified degrees of freedom, and there is a strong non-local quantum correlation between them which can not be reproduced by any classical correlation, like the violation different kinds of Bell's inequalities\cite{horodecki2009quantum,brunner2014bell,clauser1969proposed,collins2002bell}. However, there are some similarities between the nonlocality of quantum entanglement and that of wave function, such as the equivalence between the measurement-induced wave function collapse and the non-signaling theory of an entangled two-part system. Multi-particle entanglement systems sometimes exhibit single-particle behavior, such as the uncertainty relationship of quantum correlations of an EPR-like two-photon state\cite{jin2018time,li2018manipulating}. Conversely, a single particle can also exhibit entanglement characteristics in some situations\cite{tan1991nonlocality}.

\par  It is found that when a quantum particle is incident on a beam splitter, the delocalized quantum state after the beam splitter can be formulated as the entangled state between the vacuum state and the single-particle state\cite{tan1991nonlocality,van2005single,van2006reply}. It is an entanglement of particle nature in Fock space. This novel quantum state has attracted anormous attention, not only because of its unique physical properties, but also due to its potential role in the field of quantum information\cite{wildfeuer2007strong,sangouard2007long,yin2008entanglement,sangouard2008purification,salart2010purification,wildfeuer2008strong,salart2010purification,guerreiro2016demonstration,di2009entanglement,brask2013testing}. The most widely used ways to test single-photon entanglement is based on homodyne detection, through which the density matrix and quantum correlated Wigner distributions have be constructed to establish different types of Bell's inequality\cite{banaszek1999testing,babichev2004homodyne,d2006tomographic,morin2013witnessing,fuwa2015experimental,ho2014witnessing,johansen1996bell,Torlai2013violation}. It should be noted that the conventional Bell's inequality, a convincing way to test quantum correlation in one specific base, is formed by measuring the joint probability in its conjugate space\cite{collins2002bell,li2018bell}. Thus the single-particle entanglement provides a good platform for studying wave-particle duality of a quantum particle.

\par In this paper, we present a theoretical study on single-photon entanglement by Bell's inequality construction via its wave-behavior measurement. It is a two dimensional entanglement, where the bases in Fock space are vacuum state and single photon state, respectively. In contrast to particle-number state,  the wave state is defined which is a coherent superposition between the vacuum state and the single-photon state. The wave state detection is accomplished by interference with another reference wave state, the wave behavior can be detected by continuously changing their relative phase difference. The joint probability is obtained by the wave-state measurement by two far separated parts, called Alice and Bob, where the reference wave state is replaced by a weak coherent state. Finally, the violation of the Bell's inequality is tested based on joint wave state measurement.

\section{Theory}

\begin{figure}[h!]
\centering\includegraphics[width=8cm]{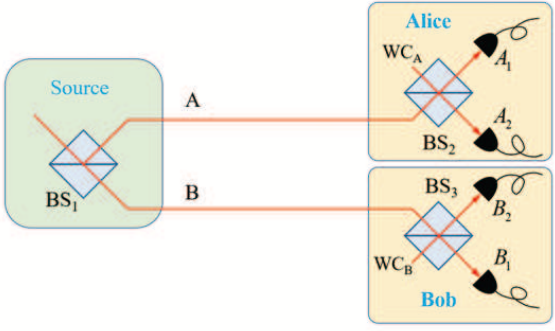}
\caption{Generation and test of single-photon entanglement. Single-photon entanglement is generated by injecting a photon into a beam splitter, in which the path modes $A$ and $B$ are entangled in Fock space between vacuum state and single-photon state. Joint probability distribution in wave space is obtained by interfering the reference weak coherent states with the wave states by Alice and Bob.}
\end{figure}

\par The generation of single-photon entanglement and the wave state measurement is illustrated in Fig. 1. A photon is incident on the first beam splitter, the two output modes A and B are entangled in Fock space
\begin{equation}
\left | \Psi_{A,B} \right \rangle= \frac{\sqrt{2}}{2}e^{i\varphi}\left[ e^{i\phi}\left | 1 \right \rangle_{A} \left | 0 \right \rangle_{B}+\left | 0 \right \rangle_{A} \left | 1 \right \rangle_{B} \right],
\end{equation}
where $\left | 0 \right \rangle$ and $\left | 1 \right \rangle$ represent vacuum state and single-photon state, $e^{i\phi}$ is the accumulated phase difference between the two arms which may be caused by half-wave loss due to reflection or different propagation path lengths, $e^{i\varphi}$ is the global phase which is always omitted. It is known that the degree of entanglement is invariable through a Fourier (unitary) transformation, and the entanglement has inverse forms in conjugate spaces. 
\par Through two-dimensional Fourier transformation, the expression of the above single-photon entanglement in wave space is
\begin{equation}
\begin{split}
\left | \Psi_{A,B} \right \rangle=& \frac{\sqrt{2}}{2} e^{i\left( \phi-\alpha \right)} [ \left | \alpha_{A} \right \rangle_{w} \left | \left(\alpha-\phi\right)_{B} \right \rangle_{w}\\
&-\left | \left(\alpha+ \pi \right )_{A} \right \rangle_{w} \left | \left( \alpha-\phi+\pi \right )_{B} \right \rangle_{w}  ],
\end{split}
\end{equation}
where the subscript $w$ means wave state, $\alpha$ is an arbitrary real number in the range of 0 to $\pi$, the two orthogonal bases in the wave space can be expressed as
\begin{equation}
\begin{split}
\left | \alpha \right \rangle_{w}=&\frac{\sqrt{2}}{2} \left [ \left | 0 \right \rangle+e^{i\alpha}\left | 1 \right \rangle \right ],\\
\left | \alpha+\pi \right \rangle_{w}=&\frac{\sqrt{2}}{2} \left [ \left | 0 \right \rangle-e^{i\alpha}\left | 1 \right \rangle \right ].
\end{split}
\end{equation}
In wave space, we use the phase value $\alpha$ to characterize a wave state. Now it is can be seen that Eqs. (2) is a diagonal form of single-photon entanglement in wave space in which the quantum correlation between the modes of Alice and Bob has a inverse form compared with that in Eqs. (1). This is a general feature of quantum correlation in conjugate spaces\cite{li2018bell,jin2018time}.

\par The particle behavior of a single-photon state is revealed by the firing of the single-photon detector, while the wave behavior of the wave state is manifested as the interference fringes when varying the phase difference between the two interference arms. Here, we put forward a method to detect the wave state by interfering it with a reference wave state, which is very similar to the quarture measurement in homodyne detection. As shown in Fig. 2, the overlapping of the wave state $\left | \alpha_{a} \right \rangle_{w}$ with a reference wave state $\left | \beta_{b} \right \rangle_{w}$ on a beam splitter is
\begin{equation}
\begin{split}
\left | \alpha_{a} \right \rangle_{w} \left | \beta_{b} \right \rangle_{w}=&\frac{1}{2}\left[ \left |0_{a} \right\rangle+e^{i\alpha}\left |1_{a} \right\rangle \right]\left[ \left |0_{b} \right\rangle+e^{i\beta}\left |1_{b} \right\rangle \right]\\
=&\frac{1}{2}  \left| 0 \right \rangle +\frac{\sqrt{2}}{4} \left ( i e^{i\alpha} +e^{i\beta} \right ) \left |1_{c} \right\rangle \left |0_{d} \right\rangle\\
& + \frac{\sqrt{2}}{4} \left (  e^{i\alpha} +i e^{i\beta} \right ) \left |0_{c} \right\rangle \left |1_{d} \right\rangle\\
&+ i\frac{1}{4}e^{i\left( \alpha+\beta \right)}\left [ \left | 2_{c} \right \rangle\left | 0_{d} \right \rangle+\left | 0_{c} \right \rangle\left | 2_{d} \right \rangle  \right ].
\end{split}
\end{equation}

\begin{figure}[h!]
\centering\includegraphics[width=3.5cm]{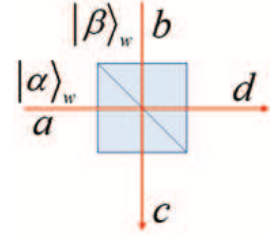}
\caption{Wave state measurement. The measurement of a wave state is performed by overlapping a reference wave state with the detected state in a beam splitter.}
\end{figure}

From Eqs. (4), we can see that the photon state behind the beam splitter consists of four parts. The first part is the vacuum state with a probability of $p\left( 0 \right)=\frac{1}{4}$; the second and the third parts are responsible for the wave behavior in which only one photon is detected in the $c$ and $d$ ports, and the probability for these two parts are $p\left( 1_{c},0_{d} \right)=\frac{1}{4}\left[ 1- \sin \left( \alpha-\beta \right) \right]$ and $p\left( 0_{c},1_{d} \right)=\frac{1}{4}\left[ 1+ \sin \left( \alpha-\beta \right) \right]$, respectively; the last part is attributed to two-photon interference which is insensitive to the phase difference. By varying the relative phase difference between the wave state and the reference state, an interference fringe can be detected, so the state has wave behavior. The fourth part of detection comes from two-photon interference, which is a kind of particle behavior interference and can be regarded as background noise.

\par The detection probability for a wave state $\left | \beta \right \rangle_{w}$ depends on the phase difference between reference wave state $\left | \alpha \right \rangle_{w}$ and detection wave state $\left | \beta \right \rangle_{w}$. As for the $c$-port, by fixing the value of $\alpha$, for $\beta=\alpha+\frac{\pi}{2}$, the detection probability of the wave state $\left | \beta \right \rangle_{w}$ is $\frac{1}{2}$, and for $\beta=\alpha-\frac{\pi}{2}$, the detection probability of the wave state $\left | \beta \right \rangle_{w}$ is $0$. However, the detection on port $d$ is completely opposite. Here you may wonder why the maximum probability of successful detection of a wave state is only $\frac{1}{2}$. This is because a successful detection of a quantum state depends on the response of the single-photon detector, which is actually a projection measurement in Fock space. According to the definition of the wave state in Eqs. (3), its detection probability on a single-photon detector is $\frac{1}{2}$, which can explain that the range of detection probability of wave state is 0 to 0.5. 

\par However, such wave state does not exist in nature, nor is there any report about the generation of the wave states. Actually, it is not easy to generate a coherent superposition of vacuum state and single-particle state directly. The state that mostly resembles a wave state is the weak coherent (WC) state which is generated by attenuating the intensity of a laser beam to well below that of a single photon. In this case, the WC state in the Fock representation can be written as
\begin{equation}
\left | \gamma e^{i \alpha} \right \rangle_{c}=\exp\left ( -\frac{1}{2} \left| \gamma \right |^{2} \right) \sum_{n=0}^{\infty} \frac{\gamma ^{n} e^{i n \alpha}}{\sqrt{n!}} \left | n \right \rangle,
\end{equation}
where the subscript $c$ represents coherent state, $\gamma$ is a positive real number, $\left | \gamma \right |^{2}$ is the average photon number in the coherent state, $\varphi$ is the phase carried by the state. If the condition $\left | \gamma \right |^{2}\ll 1$ is satisfied, the weak coherent state can be approximated as
\begin{equation}
\left | \gamma e^{i \alpha} \right \rangle_{c}\approx \left | 0 \right \rangle +\gamma e^{i \alpha} \left | 1 \right \rangle +O\left( \gamma \right) \left | n \geq 2 \right \rangle.
\end{equation}
In the power series expansion of the WC state in Fock space, only the vacuum state and the single-photon state is retained, all the high order infinitesimal terms are represented by $O\left ( \gamma \right ) \left | n \geq 2 \right \rangle$. Interference between two wave states approximated by WC states is the most fundamental physics in phase-matching quantum key distribution\cite{lucamarini2018overcoming,ma2018phase}.

\par Equipped with the technique of wave state detection given above, we derive the joint probability distribution through the use of wave state measurement for single-photon entanglement. The reference WC states possessed by Alice and Bob are $\left | \gamma_{A} e^{i \alpha} \right \rangle_{c}$ and $\left | \gamma_{B} e^{i \beta} \right \rangle_{c}$, respectively. To maximize the resolution of the interference pattern, we first fix $\gamma_{A}=\gamma_{B}=\gamma$. According to the definition of wave state measurement in Fig. 2 and the detection devices in Fig. 1, for the reference wave states of $\left | \gamma e^{i \alpha} \right \rangle_{c}$ and $\left | \gamma e^{i \beta} \right \rangle_{c}$ used by Alice and Bob, the wave states detected in single photon entanglement are $\left | \alpha+\frac{\pi}{2} \right \rangle$ and $\left | \beta+\frac{\pi}{2} \right \rangle$ at ports $A_{1}$ and $B_{1}$, and $\left | \alpha-\frac{\pi}{2} \right \rangle$ and $\left | \beta-\frac{\pi}{2} \right \rangle$ at ports $A_{2}$ and $B_{2}$ respectively. By overlapping two reference WC states with the single-photon entangled state $\left | \Psi_{A,B} \right \rangle$ in the beam splitters by two spatially separated parts, Alice and Bob, the joint state $\left | \Psi \right \rangle$ is
\begin{equation}
\begin{split}
\left | \Psi \right \rangle=&\left | \gamma e^{i \alpha} \right \rangle_{c}\left | \gamma e^{i \beta} \right \rangle_{c} \left | \Psi_{A,B} \right \rangle\\
\approx&\frac{1}{2}\left [ ie^{i\phi }\left [ \left | 1_{A_{1}} \right \rangle+i\left | 1_{A_{2}} \right \rangle \right ]+\left [ \left | 1_{B_{1}} \right \rangle+i\left | 1_{B_{2}} \right \rangle \right ] \right ]\\
+&\frac{\sqrt{2}}{4}\gamma e^{i\left (\phi+\beta  \right ) } \left [ \left |1_{A_{1}}\right \rangle+i\left |1_{A_{2}}\right \rangle \right ]\left [ i\left |1_{B_{1}}\right \rangle+\left |1_{B_{2}}\right \rangle \right ]\\
+&\frac{\sqrt{2}}{4}\gamma e^{i\alpha }\left [i \left |1_{A_{1}}\right \rangle+\left |1_{A_{2}}\right \rangle \right ] \left [ \left |1_{B_{1}}\right \rangle+i\left |1_{B_{2}}\right \rangle \right ]\\
+&\frac{\sqrt{2}}{4}\gamma e^{i\beta }\left [ \left |1_{B_{1}}\right \rangle+i\left |1_{B_{2}}\right \rangle \right ] \left [ i\left |1_{B_{1}}\right \rangle+\left |1_{B_{2}}\right \rangle \right ]\\
+&\frac{\sqrt{2}}{4}\gamma e^{i\left (\alpha+\phi \right )}\left [ \left |1_{A_{1}}\right \rangle+i\left |1_{A_{2}}\right \rangle \right ] \left [ i\left |1_{A_{1}}\right \rangle+\left |1_{A_{2}}\right \rangle \right ]\\
+& O\left( \gamma \right ) \left | \text{other} \right \rangle.
\end{split}
\end{equation}
Here we mainly focus on the quantum correlation between A and B, so only the coincidence counts between them are considered. According to Eqs. (4), a part of the counting in the wave state detection comes from two-photon interference, which is insensitive to phase difference. By setting $\alpha'=\alpha+\frac{\pi}{2}$ and $\beta'=\beta+\frac{\pi}{2}$, the joint probabilities between wave states detected at the four single-photon detectors can be approximated to
\begin{equation}
\begin{split}
p\left(A_{1},B_{1} \right)=&\frac{\gamma^{2}}{4}\left[1+\cos \left( \alpha'-\beta'-\phi \right ) \right]+\frac{\gamma^{4}}{4},\\
p\left(A_{1},B_{2} \right)=&\frac{\gamma^{2}}{4}\left[1-\cos \left( \alpha'-\beta'-\phi \right ) \right]+\frac{\gamma^{4}}{4},\\
p\left(A_{2},B_{1} \right)=&\frac{\gamma^{2}}{4}\left[1-\cos \left( \alpha'-\beta'-\phi \right ) \right]+\frac{\gamma^{4}}{4},\\
p\left(A_{2},B_{2} \right)=&\frac{\gamma^{2}}{4}\left[1+\cos \left( \alpha'-\beta'-\phi \right ) \right]+\frac{\gamma^{4}}{4}.
\end{split}
\end{equation}
From Eqs. (8), we can see that the joint probabilities contain two parts. The first part comes from the phase-sensitive interference, which represents the wave-like correlation between Alice and Bob. The second part can be viewed as a  phase-insensitive interference, which represents the particle-like correlation between Alice and Bob. For the weak coherent state as the reference wave state under the condition of $\left | \gamma \right |^{2}\ll 1$, the second part of the coincidence count can be ignored. In Eqs. (8), the phase $\phi$ arises from the optical path difference between two photon state modes $A$ and $B$ propagating from the light source to Alice and Bob, where its value can be considered as fixed. The coincidence counts between the single-photon detectors owned by Alice and Bob is a cosine function with respect to the phase difference of the wave states between them. Also the visibility of the joint probability curve reaches the maximum value of 1 under the condition that the average photon number of the WC state $\gamma^{2}$ is far below than 1. Thus the derivation of joint probability in wave space gives the wave-particle duality in one-photon entanglement accurately from another aspect.

\par By removing the single-photon counts that do not contribute to the coincidence counts, from Eqs. (8), it is can be seen that the count rate of each single-photon detector, which is the marginal probability of $p\left( A_{i},B_{j} \right )$ with $i,j=1,2$, is equal to a constant
\begin{equation}
p\left( A_{1} \right )=p\left( A_{2} \right )=p\left( B_{1} \right )=p\left( B_{2} \right )=\frac{\gamma^{2}}{4}.
\end{equation}
Thus it is can be seen that the correlation between the wave state measurements of the far separated two parts can be described as first order coherence, a common feature of quantum correlation with maximum entanglement. The completely different behaviors between the single count rate and the coincidence count rate also demonstrate the quantum nonlocality of a single photon state.

\par To give a conclusive conjugation between wave behavior and particle behavior, we need to verify the existence of single photon entanglement of particle behavior by using the joint probability derived from wave space. To rule out possibility that measurement results can instead be explained by local hidden variable (LHV) theories, a violation of the Bell's inequality based on the joint probability in Eqs. (9) should be tested. The most commonly used CHSH-type Bell's inequality for two-dimensional entanglement can be written as\cite{clauser1969proposed,yarnall2007experimental}
\begin{equation}
\begin{split}
S=& | E\left ( \alpha' _{1},\beta' _{1} \right )+E\left ( \alpha' _{1},\beta' _{2} \right )\\
&+E\left ( \alpha' _{2},\beta' _{1} \right )-E\left ( \alpha' _{2},\beta' _{2} \right )  |\leq 2,
\end{split}
\end{equation}
where $\alpha'_{1(2)}$ and $\beta'_{1(2)}$ represent the wave states in the single-photon entanglement detected by Alice and Bob, respectively; $E\left ( \alpha' _{i},\beta' _{j} \right )$ is the correlation function which can be written as
\begin{equation}
\begin{split}
E\left ( \alpha' _{i},\beta' _{j} \right )=&P\left(A_{1},B_{1} \right)+P\left(A_{2},B_{2} \right)\\
&-P\left(A_{1},B_{2} \right)-P\left(A_{2},B_{1} \right).
\end{split}
\end{equation}
Here $P\left(A_{i},B_{j} \right)$ is the normalized joint probability
\begin{equation}
P\left(A_{i},B_{j} \right)=\frac{p\left(A_{i},B_{j} \right)}{\sum_{l=1,m=1}^{2,2}p\left(A_{m},B_{n}\right)},
\end{equation}
where the subscripts $i,j,m,n$ represent the indices of the single-photon detectors. If we assume the phase difference between the modes distributed to Alice and Bob $\phi=0$, by setting $\alpha'_{1}=0$, $\alpha'_{2}=\frac{\pi}{2}$ and $\beta'_{1}=\frac{\pi}{4}$, $\beta'_{2}=-\frac{\pi}{4}$, a result of $S=2\sqrt{2}$ can be obtained. The violation of Bell's inequality in Eqs. (8) demonstrate not only the existence of delocalized single-photon entanglement but also the wave-particle duality in single-photon entanglement. 

\section{Discussion and Conclusion}

\par Now, we would like to point out the difference between the methods of testing Bell's inequality, like phase correlation\cite{banaszek1999testing}, Wigner distribution correlation\cite{babichev2004homodyne,fuwa2015experimental}, quadrature correlation\cite{morin2013witnessing,ho2014witnessing} and wave state correlation in our paper. The similarity of these methods is that single-photon quantum correlation is detected by interfering the path modes with local coherent states. Here we only focus on the different manifestations of quantum correlation, regardless of the specific forms of Bell inequality in different methods. Because wave states are obtained by the Fourier transform of the particle states characterized by the phase gradient, thus the quantum correlation between far separated wave states can be well exhibited by the quantum correlation between the phases of two local oscillators\cite{banaszek1999testing}. In fact, this phenomenon can not be interpreted as the quantum correlation between the phases of the two far separated local oscillators, as they are independent of each other. It is the quantum correlation between the wave states that leads to this novel phenomenon. The phase measurement of local oscillator can also be represented by quadrature measurement of electromagnetic field in homodyne detection\cite{vogel1989determination}, thus another representation of phase correlation can also be represented by quadrature correlation\cite{morin2013witnessing}. Furthermore, the Wigner distribution of a quantum state can be measured by the quadrature measurement of a homodyne detection\cite{cahill1969density,cahill1969ordered,yuen1980optical,yuen1983noise,schumaker1984noise,yurke1987measurement}, thus the quadrature correlation will lead to the Wigner distribution correlation as well as correlation of the density of states\cite{babichev2004homodyne}. The difference between these methods lie in that entanglement is carried by different types of carriers. Through wave state measurement proposed in this paper, we can better understand the relationship between wave entanglement and particle entanglement in a single photon entanglement.

\par Next, we also want to give a simple discussion about the wave-particle duality in single-photon entanglement. Bohr's complementary principle states that a full description of a quantum entity should take into account of both the wave-behavior and the particle-behavior, i.e. the wave-particle duality. The particle behavior and wave-behavior can be selectively present depending on the experimental configuration\cite{kaiser2012entanglement,rab2017entanglement}. The simplest illustration of a photon is the photoelectric effect where electrons emit from the host material hit with a light beam with enough frequency. The most direct application based on this attribute is the single-photon detector, a basic element in the field of quantum information. The particle behavior is always discussed in the Fock space, i.e. the particle number space within which the quantum states are called particle number states. The wave-behavior can be evidenced in the Young-type double-slit experiments where the distribution of the detection probability for a single-photon form an interference fringe\cite{ma2016delayed,walborn2002double,tang2012realization}. In our work, however, wave and particle have more specific meanings. They are a pair of conjugated observable quantities. This is reflected in the fact that single-photon entanglement can be described in both particle-state and wave-state. Therefore, their eigen-projection measurements are incompatible with each other. It also conforms to the complementary principle that the particle-behavior and wave-behavior can not be detected at the same time. At this point, our results point out the equivalence between complementarity principle and Heisenberg uncertainty principle.

\par In summary, we have proposed a method to demonstrate the existence of a delocalized single-photon entanglement by constructing the CHSH-type Bell's inequality based on joint measurement of wave states. By Fourier transform, the entanglement of single photon in Fock space is transformed into the wave space, which has the opposite diagonalization form. Our results show that the entanglement is not only restricted to multi-particle systems and the carrier of the entanglement is not necessarily a single particle. The Bell's inequality violation of single-photon entanglement points out the intrinsic relationship between the nonlocality of wave function and that of quantum entanglement. The introduction of wave state adds a new degree of freedom to the quantum system, which can be used as a new information carrier in the field of quantum information. The definition and detection method of a wave state presented in this paper can be extended to arbitrary quantum single particle systems.

\acknowledgments
This work is supported by Young fund of Jiangsu Natural Science Foundation of China (SJ216025), National fund incubation project (NY217024), Scientific Research Foundation of Nanjing University of Posts and Telecommunications (NY215034), the National Natural Science Foundation of China (No. 61475075), the open subject of National Laboratory of Solid State Microstructures of Nanjing University (M31021).

\bibliography{reference}
\end{document}